\documentstyle[12pt,aaspp4]{article}
\newcommand{\ifms}[1]{}
\newcommand{\ifpp}[1]{#1}
\newcommand{\degree}{\ifmmode {^{\circ}} \else {$^{\circ}$} \fi}
\newcommand{\degrees}{\ifmmode {^{\circ}} \else {$^{\circ}$} \fi}

\newcommand{\unit}[1]{\ifmmode {\rm\ #1\,} \else {$\rm #1$} \fi}
\newcommand{\quarter}{\ifmmode {\frac{1}{4}} \else {$\frac{1}{4}$} \fi}
\newcommand{\angstrom}{\unit{\AA}}

\newcommand{\etal}{{et al.~}}

\newcommand{\km}{\unit{km}}

\newcommand{\tten}[1]{\ifmmode {\times 10^{#1}} \else {$\times 10^{#1}$} \fi}
\newcommand{\tentothe}[1]{\ifmmode {10^{#1}} \else {$10^{#1}$} \fi}

\newcommand{\microsec}{\unit{\mu s}}

\newcommand{\pu}{\unit{ph\ s^{-1}}\unit{cm^{-2}\ str^{-1}}}

\newcommand{\doublet}{\ifmmode {\lambda\lambda} \else {$\lambda\lambda$} \fi}
\newcommand{\singlet}{\ifmmode {\lambda} \else {$\lambda$} \fi}

\newcommand{\percmsqr}{\unit{cm^{-2}}}

\newcommand{\sciama}{\tten{-16} \unit{s^{-1} cm^{-3}}}
\newcommand{\mgf}{\unit{MgF_2}}
\ifpp{\input psfig}
\begin{document}

\lefthead{Bowyer \etal}
\righthead{Evidence Against Sciama Model of Massive Neutrinos}

\title{Evidence Against the Sciama Model of Radiative Decay of Massive   
Neutrinos
\footnote{Based on the development and utilization of the Espectr\'ografo
Ultravioleta de Radiaci\'on Difusa, a collaboration of the Spanish Instituto 
Nacional de Tecnica Aeroespacial and the Center for EUV Astrophysics, 
University of California, Berkeley}}

\author{Stuart Bowyer, Eric J. Korpela, Jerry Edelstein, and Michael Lampton} 
\affil{Space Sciences Laboratory, University of California, Berkeley, CA 
  94720-7450}

\author{Carmen Morales, Juan P\'erez-Mercader, Jos\'e F. G\'omez
\altaffilmark{2}, 
and Joaqu\'{\i}n Trapero\altaffilmark{3}}
\affil{Laboratorio Astrof\'{\i}sica Espacial y F\'{\i}sica
Fundamental, 
INTA.  Apdo.
Correos 50727, 28080 Madrid, Spain}

\altaffiltext{2}{Also at Instituto de Astrof\`{\i}sica de
Andaluc\'{\i}a, CSIC, Apdo Correos 3004, E-18080, Granada,
Spain}
\altaffiltext{3}{Present address: Universidad SEK, Cardenal Z\'u\~niga
S/n, Segovia, Spain}

\begin{abstract}	
We report on spectral observations of the night sky in the band around 900 \angstrom 
where 
the emission line in the Sciama model of radiatively decaying massive 
neutrinos would be present. The data were obtained with a high resolution, 
high sensitivity spectrometer flown on the Spanish MINISAT satellite. The 
observed emission is far less intense than that expected in the Sciama model.
\end{abstract}

\keywords{cosmology: diffuse radiation --- elementary particles --- 
ultraviolet: general}

\section{Introduction}

Relic neutrinos, if massive, could contribute significantly to the density of the 
universe, and if appropriately concentrated, could explain puzzling 
characteristics of luminous matter in galaxies.  Melott (1984) suggested that
if these particles were
radiatively decaying, they could be responsible for the sharp hydrogen
ionization edges seen in many galaxies and that this decay would not violate
existing observational data if the decay energy was somewhat greater than 13
eV and the lifetime for decay was about 10$^{24}$ s.  In a subsequent paper,
Melott \etal 1988 showed this idea was consistent with observations of
star formation, galaxy formation and morphology, and other phenomena.
Subsequently, Sciama and collaborators in an extensive set of papers (Sciama 1990, 
1993, 1995, 
1997a, 1997b, 1998, Sciama \etal 1993) showed that if the decay lifetime was an 
order of magnitude less than that suggested by Melott his theory 
could explain a large number of otherwise puzzling astronomical 
phenomena, including the ionization state of the intergalactic medium and 
the anomalous ionization of  the interstellar medium (ISM) in our own 
Milky Way Galaxy.  Although massive neutrinos cannot be contemplated within the
framework of the standard model of particle physics, they can be accommodated in the supersymmetric
extensions of the standard model, especially if R-parity is broken (cf. Gato \etal 1985, Bowyer \etal, 1995). Recent observational and experimental results suggest they do, 
in fact, have mass (Fukuda \etal 1998a,b  and Athanassopoulos
\etal 1998a,b).

A number of searches have been made for evidence of radiatively decaying massive 
neutrinos in clusters of galaxies. Davidsen \etal (1991) severely constrained 
the parameter space available for these particles through 
observations of the cluster of galaxies Abell  665, and Fabian \etal 1991 obtained similar results from a study of the cluster of galaxies surrounding 
the quasar 3C263.  However, Sciama \etal (1993) and 
Bowyer \etal (1995)  have shown that these observations do not rule out the 
Sciama scenario.

An all-pervading neutrino flux in the Galaxy at a wavelength near the 
ionization limit of hydrogen would be difficult to observe because of 
absorption by the ISM.  However, Bowyer \etal (1995) pointed out that this 
flux would be observable from Earth orbit in several well-defined directions 
where the density of the ISM is extremely low.  An observational 
complexity which could complicate these measurements is emission from an 
upper atmosphere oxygen recombination feature at 911 \angstrom (Chakrabarti
\etal 1983). 

In this paper we report results of spectral observations made in the region $\leq $
912 \angstrom where the radiation in the Sciama scenario would be present, and compare 
the data obtained with the flux expected.  

\section{Observations}

The observations were made with an extreme ultraviolet spectrometer 
covering the band-pass from 350-1100 \angstrom which was specifically designed for 
studies of diffuse emission.  The instrument (the Espectr\'ografo Ultravioleta 
extremo para la Radiaci\'on Difusa, EURD) is capable of
providing measurements of the diffuse UV background which are more than 100 times more sensitive 
than existing measurements in this band-pass, with a spectral resolution of 
about 6\angstrom.
The instrument is described in detail by Bowyer \etal 1997.

The instrument was flown onboard the Spanish MINISAT-01 satellite launched on April 
21, 1997.  The spacecraft is in a retrograde orbit with an inclination of 151 $^\circ$ and is at an altitude of 575 \km. The spectrometer continuously 
views the anti-Sun direction.  Details of the spacecraft and the EURD 
observational parameters are provided in Morales \etal 1996.  

We examined EURD data in the 890 to 915 \angstrom bandpass in an attempt to detect the 
emission which would be present if the Sciama scenario was operative.
Data from the spectrometer were typically collected over the entire night-time 
portion of the orbit. Higher count rates are always experienced at spacecraft sunrise 
and sunset due to geocoronal effects, but deep night intensities are 
typically constant and low.  For the search for radiation from  the Sciama 
scenario, we sorted the data to exclude all sunrise and sunset data and all other
data associated with high backgrounds. Given the low in-flight counting rate and 
the absolute fixed electronics dead time of 100 \microsec per photon, dead time 
corrections were about 1 \% and were therefore ignored.

The EURD spectrograph employs a number of vetoes to reduce unwanted background 
and to permit evaluation of those background events which cannot be otherwise eliminated (Bowyer \etal l997).  The detector is surrounded by an 
anti-coincidence shield and
all counts triggering this shield (about 20 percent) are rejected.  Remaining
internal background components include charged particles that are missed
by the anti-coincidence system, Compton scattered $\gamma$-rays, and radioactivity
within the detector and in the spacecraft.  An additional background is
produced by photons scattered by the grating onto the detector. This
scattered emission is mostly a continuum arising from the wings of the zero
and first order of the hydrogen Lyman-alpha line whose peaks 
were designed to fall beyond the ends of the detector.  

The entrance aperture of the instrument has a filter wheel with three positions:
Open, Closed, and a \mgf filter.  
The Open position provides spectral data plus backgrounds. The Closed position gives 
an estimate of the internal background, and the \mgf filter position gives an estimate 
 of the scattered radiation. Observations were carried out sequentially with each of 
these apertures; the complete cycle time was 90 s.

We corrected the deep night spectral data for backgrounds using the \mgf 
and Closed apertures. We summed the background corrected 
data in the 890 to 
915 \angstrom band as a function of time. We included data to 915 \angstrom to 
assure all 
counts shortward of 912 \angstrom were included in the sample given the spectral 
resolution of 
the instrument. In some neutrino decay scenarios, two lines will be produced 
whose relative intensities are uncertain. However, the sum of both of these 
lines is the key parameter to be measured, and in the Sciama scenario these 
lines will be separated by  0.2 eV, or 13 \angstrom  at 900 \angstrom. 
Hence the flux from 
both these lines will be included in the data reported here. For this study we utilized data
obtained between 18 June 1997 and 29 June 1998.  Data were 
regularly obtained over most of this period, with occasional gaps because
of spacecraft problems or instrument shutdowns.  Data were summed over 10 day
intervals, providing typically about 3500 counts, to obtain good counting 
statistics.
\ifpp{
\begin{figure}[t]
\begin{center}
\ \psfig{file=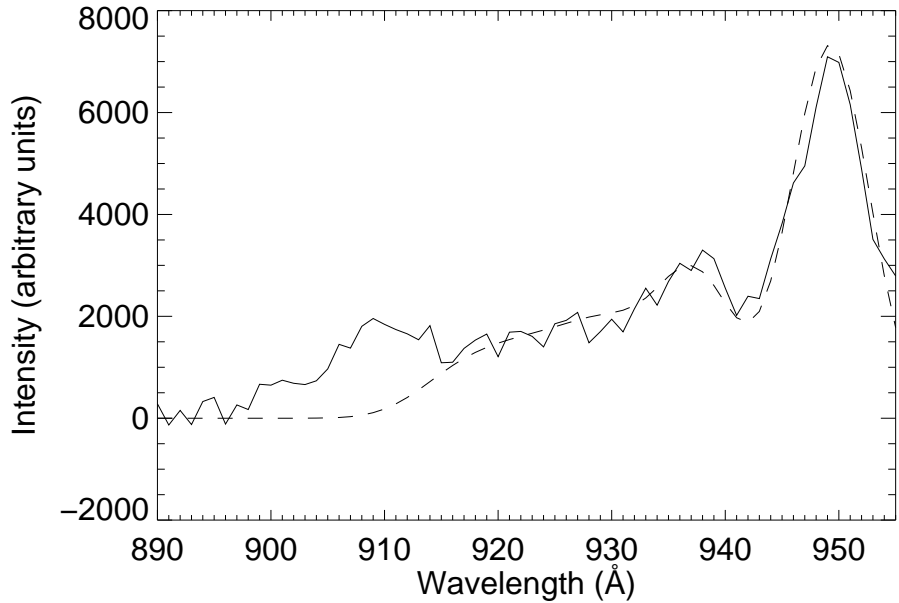}
\end{center}
\caption{\small The observed spectrum of the region around 911 \angstrom where r
adiation 
in  the Sciama scenario is expected.  The solid line shows a straight line 
interpolation of the spectrum from about 60 hours of shutter open 
observations during a period when emission from the oxygen recombination 
feature was more pronounced.  The dashed line shows the expected 
hydrogen Lyman series lines with an electron temperature of 0.4eV folded 
with the instrument resolution and fit to the average intensity of the Lyman 
series lines during this period.}
\end{figure}
}

Unfortunately in regards to our search for the Sciama line, oxygen 
recombination radiation was substantial at the altitude of
the MINISAT satellite even in the anti-Sun view direction.
A spectrum of the radiation detected around 912 \angstrom 
is shown in 
Fig.~1.  This spectrum shows a profile that is consistent with 
the line shape obtained by 
Feldman \etal 1992 given the resolution of this instrument.  Just longward 
of 912 \angstrom the spectrum is dominated by the Lyman series lines of 
geocoronal hydrogen (L\'opez-Moreno \etal  l998).  Both the oxygen recombination 
feature and the Lyman series of hydrogen vary in time; the data shown in 
Fig.~1 are from a period when the oxygen recombination radiation was 
more pronounced.   

We determined the EURD counts-to-flux conversion factor in
the region around 800 \angstrom
using an in-flight calibration strategy based on simultaneous EUV observations of the Moon with EUVE and 
EURD (Flynn \etal 1998), and, longward of 912 \angstrom, to fits to stellar spectra 
(Morales \etal, in progress).  It is estimated that this calibration 
is good to 
$\pm$ 20\% in the band around 912 \angstrom because of the quality of the fit to 
stellar spectra.  This in-flight calibration yields a conversion of 6.5 \tten{4}
\unit{ph\,cm^{-2}\,str^{-1}} per count at 912 \angstrom; this is within a factor of 
three of the 
preflight calibration (Bowyer \etal l997). This difference is easily
understood as being the result of the degradation of the
detector photocathode during the almost 2 year time period between the laboratory
calibration and in-orbit operation. The resulting fluxes are shown in 
Fig.~2.  These fluxes are the total fluxes obtained in this
bandpass, uncorrected for any Lyman series emission as seen
in Fig.~1.

\ifpp{
\begin{figure}[t]
\begin{center}
\ \psfig{file=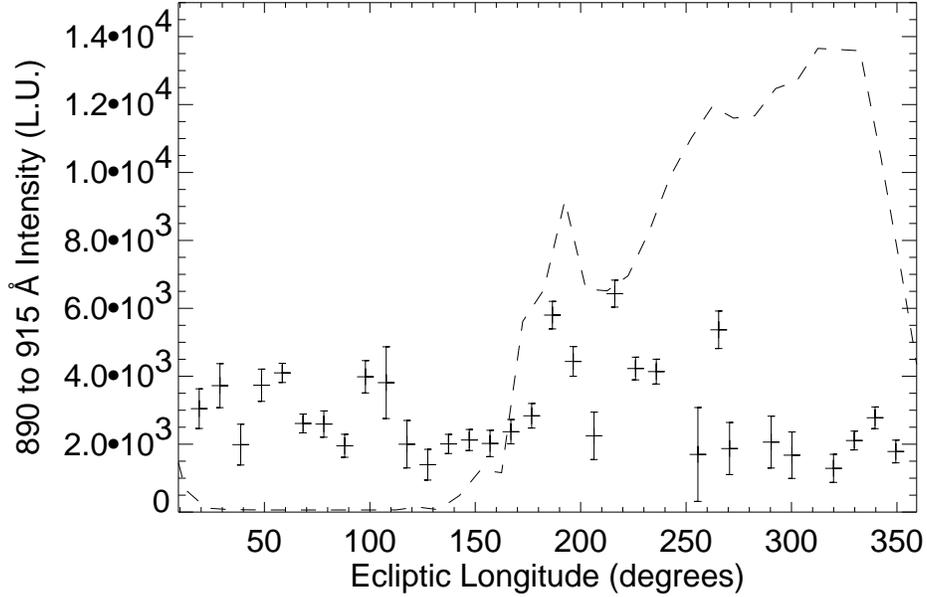}
\end{center}
\caption{\small The expected signature from the radiative decay of massive 
neutrinos as predicted by the Sciama scenario is shown as a dashed line.  This s
ignature depends upon ecliptic longitude because of absorption by the LIC and th
e size of the open region beyond.
The integrated
intensity of the observed emission from 890 to 915 \angstrom is also shown.  Eac
h data 
point is a sum of 10 days of deep night data. }
\end{figure}
}

The expected emission in the Sciama scenario can best be considered in two 
parts. The first is produced in the Local Interstellar Cloud (LIC) which surrounds the Sun; this emission is 
intermixed with absorption. The second component is emission from beyond the 
LIC which is absorbed by this cloud. 

Formally, the emission  is given by the relation:                                   
\begin{equation}
I(l) = B + {{R_{\rm prod}}\over{4\pi n_o \sigma}}\left(1-\exp\left[-n_o \sigma d_{\rm cl}(l)\right] \right) +
        {{R_{\rm prod}\left[d_e(l) - d_{\rm cl}(l)\right]}\over{4\pi}}\exp\left[- n_o  \sigma d_{\rm cl}(l)\right]
\end{equation}

where we have included a background, $B$ (which could be due to anything, but is mostly due to oxygen recombination radiation); $R_{\rm prod}$ is the photon production rate; $n_o$ is 
the density of the LIC; $\sigma$ is the effective ISM cross section for 
absorption (Rumph \etal 1994); $d_{\rm cl}$ is the distance to the cloud edge; and 
$d_e$ is the distance to the edge of the neutral free region. The symbol $l$
indicates variation with ecliptic longitude. The most recent (small) revision 
of the theory (Sciama l998) requires a photon production rate of 
$2 \pm 1 \sciama$.

We have used the model of Redfield and Linsky (1999) for data on 
the LIC.  This is a three dimensional model which 
is based on ISM absorption features in the spectra of nearby
stars obtained with HST, EUVE, and ground based telescopes.
Minimum hydrogen columns in the plane of the ecliptic in this
model are $\sim 2.5 \times 10^{16} \percmsqr$, maximum columns are $\sim 2.5\times 10^{18} \percmsqr$.

In the region beyond the LIC, Welsh \etal 1998 used high
resolution optical spectroscopy to determine the amount of
ISM sodium in the line of sight to stars within 300 pc of the Sun. They found that the ISM is essentially free of neutral gas out
to more than 70 pc in most directions. Sfeir \etal 1999 have
obtained an extensive set of sodium absorption data and have
modeled the extent of this ionized region, or Local Bubble.
We have used the  
N(H) = 1\tten{19} \percmsqr contour of their model, where the ionized
 region of the Local Bubble abruptly ends, as the limit to the region from which the 
Sciama line could be detected.
This contour is typically at 100 pc in the plane of the
ecliptic.
We have incorporated these results in Eqn.~1, and we show the expected 
emission in the plane of the ecliptic for the Sciama scenario in Fig.~2.

\section{Discussion  and Conclusions}

The geocoronal oxygen background is obvious in the data shown in Fig.~2, 
but in those view directions in which the absorption by the LIC is small
because of the Sun's location within the cloud, the flux from 
radiatively decaying neutrinos should be far more intense than the 
oxygen emission. It is obvious by inspection that the emission predicted by
the Sciama theory is not present.

We have fit our data shown in Fig. 2 to a model described by Eqn. 1 in which we
treat the background
$B$ and the photon production rate $R_{\rm prod}$ as
parameters.  Our best fit value for $B$ is 2200 \pu.  Our best fit for 
$R_{\rm prod}$ is consistent with zero and has a 95\% confidence upper limit of 
0.6 \sciama, which is one third of the production rate required by the
theory.

The EURD data appear to be completely incompatible with the Sciama 
model of radiatively decaying massive neutrinos. We believe that the only 
parameters in this study that could be challenged, in principle, are 
the conversion factor from observed EURD counts-to-flux, and the LIC 
model. In evaluating the calibration issue, we note that while the most 
accurate conversion factor can be derived from stellar spectra, this result 
requires substantial justification (to be discussed elsewhere) and is not 
necessary for this work. The EURD in-flight calibration is firmly established to 
within a factor of two through our in-flight observations of the intensity of the
geocoronal hydrogen Lyman 
lines. The counts-to-flux conversion using these in-flight results would have to be incorrect by more 
than a factor of five to reduce the predicted emission in the Sciama scenario 
to the level of 
the background shown in Fig.~2 if all the uncertainties are added in their worst directions. We can think of no way that this could 
be realized.  The other factor that could be challenged, the LIC model, 
would have to be incorrect by a factor of twenty to reduce the observed flux 
to the level of the background shown in Fig.~2. This possibility is considered 
to be extremely unlikely (J. Linsky, private communication).  
  
Although we believe our data rule out the Sciama model of
radiatively decaying neutrinos, we note that we cannot
exclude the earlier model of Melott with its longer
lifetime.  In this respect, it is intriguing to note that we
do observe a faint line at $\sim 710 \AA$ in long
integrations with the EURD instrument which we have not been able to
identify as either an upper atmospheric airglow line or as
emission from the interstellar medium. (Bowyer \etal, in
progress)

\section{Acknowledgements}

We wish to acknowledge many useful discussions with Dennis Sciama. We 
thank Jeff Linsky and Seth Redfield for access to their model before 
publication, and Seth Redfield for help in utilizing this model.
We than Daphne Sfeir for access to her model before
publication and her help in utilizing this model. The authors 
wish to thank J. Cobb for devising and implementing complex data 
processing programs which convert the spacecraft data to forms amenable to 
scientific analysis.  Partial support for the development of the EURD 
instrument was provided by NASA grant NGR 05-003-450 and INTA grant IGE
490056. 

When the NASA funds 
were withdrawn from the instrument development at Berkeley
 by Ed Weiler, the instrument was completed with 
funds provided by S. Bowyer.  
The UCB analysis and interpretation is carried out through the 
volunteer efforts of S. Bowyer, J. Cobb, J. Edelstein, E. Korpela, and M. 
Lampton.  
The work by C. Morales and J. Trapero is supported in part by DGCYT grant PB94-0007.  
J.F. G\'omez is supported in part by DGCYT grant PB95-0066 and
Junta de Andalucia (Spain). J. P\'erez-Mercader is supported by 
funds provided by the Spanish ministries of Education and
Defense.

\clearpage

\ifms{

\figcaption[fig1.eps]{The observed spectrum of the region around 911 \angstrom where radiation 
in  the Sciama scenario is expected.  The solid line shows a straight line 
interpolation of the spectrum from about 60 hours of shutter-open 
observations during a period when emission from the oxygen recombination 
feature was more pronounced.  The dashed line shows the expected 
hydrogen Lyman series lines with an electron temperature of 0.4eV folded 
with the instrument resolution and fit to the average intensity of the Lyman 
series lines during this period.}

\figcaption[fig2.eps]{The expected signature from the radiative decay of massive 
neutrinos as predicted by the Sciama scenario is shown as a dashed line.  This 
signature depends upon ecliptic longitude because of absorption by the LIC and the 
size of the open region beyond.
The integrated
intensity of the observed emission from 890 to 915 \angstrom is also shown.  Each data 
point is a sum of 10 days of deep night data. }

}

\end{document}